\documentclass[12pt,a4paper]{article}
\pdfoutput=1
\RequirePackage{amsmath}%
\RequirePackage{amsthm}%
\usepackage{amsfonts}%
\usepackage{amssymb}%
\usepackage{graphicx}%
\usepackage{url}%
\usepackage{booktabs}
\usepackage{multirow}

\renewcommand{\Re}{\mathop{\mathrm{Re}}}
\renewcommand{\Im}{\mathop{\mathrm{Im}}}

\renewcommand{\i}{\mathrm{i}}

\newcommand{\bI}{{\bf I}}

\newcommand{\bM}{{\bf M}}
\newcommand{\bN}{{\bf N}}

\newcommand{\bn}{{\bf n}}

\begin{document}
\title{Application of integral equations to simulate local fields in carbon nanotube reinforced composites}
	\author{Mohamed M S Nasser$^{\rm a}$ and El Mostafa Kalmoun$^{\rm b}$}
	
	\date{}
	\maketitle
	
	\vskip-0.8cm %
	\centerline{$^{\rm a,b}$Department of Mathematics, Statistics and Physics, Qatar University,} %
	\centerline{P.O. Box 2713, Doha, Qatar.}%
	\centerline{E-mail: $^{\rm a}$mms.nasser@qu.edu.qa, $^{\rm b}$ekalmoun@qu.edu.qa}

\begin{abstract}
We consider the steady heat conduction problem within a thermal isotropic and homogeneous square ring  composite reinforced by non-overlapping and randomly distributed carbon nanotubes (CNTs).  We treat the CNTs as rigid line inclusions and assume their temperature distribution to be fixed to an undetermined constant value along each line. 
We suppose also that the temperature distribution is known on the outer boundary and  that there is no heat flux through the inner square. The equations for the temperature distribution are governed by the two-dimensional Laplace equation with mixed Dirichlet-Neumann boundary conditions. This boundary value problem is solved using a boundary integral equation method. We demonstrate the performance of our approach through four numerical examples with small and large numbers of CNTs and different side length of the inner square.
\end{abstract}

\begin{center}
\begin{quotation}
\end{quotation}
\end{center}

\section{Introduction}
The macroscopic conductivity of polymer composites reinforced by carbon nanotubes and simulation of their stationary local fields allow to improve the rheological and processing properties \cite{nano1}. The main properties of conductive composites such as electrical, dielectric and heat conductivity (resistivity) can be evaluated by advanced computer simulation of the 2D structures. The same mathematical formalism describes the anti-plane shear fields in composites, hence, it can be applied to estimation of the elastic fields in such a composite \cite{nano2}. For definiteness, we use the heat conduction terminology in the present paper.   


The stationary heat conduction equations in a two-dimensional isotropic medium containing inclusions or voids can be expressed in the form of a mixed Dirichlet-Neumann boundary value problem in a multiply connected domain. Complex variable methods are powerful tools for solving such a problem~\cite{gak,Mit-chap,Mit-Rog,Nish,Mit-mod,Wen92,mus}. An appealing property of this problem is its invariance under conformal mappings. Consequently, if the underlying physical domain has a complicated geometry, conformal mappings can be used to map it onto a new domain of simpler geometry. One way of solving the mixed Dirichlet-Neumann boundary value problem is by reducing it to a Riemann-Hilbert problem~\cite{gak,Mit-Rog,Wen92}, which can be solved by a method of functional equations~\cite{dry,Mit-chap,Mit-Rog,Mit-mod,Ryl}. Alternatively, the Riemann-Hilbert problem can be solved using the boundary integral equation method with the generalized Neumann kernel~\cite{nas-bvp,Nas-ETNA,Nas-Gre,Nas-amc,Weg-Nas}. 

On the other hand, a hypersingular boundary integral equation was used in~\cite{Nish} to solve the steady heat conduction problem in carbon nanotubes (CNTs) composites~\cite{Zha04}. The domain considered in~\cite{Nish} is unbounded and contains no isolation region, which yields a boundary value problem with Dirichlet boundary conditions. Moreover, the authors treated only the case of CNTs that are aligned in the $x$-axis direction.

\begin{figure}[ht] %
	\centerline{
		\scalebox{0.6}{\includegraphics[trim=0 0 0 0,clip]{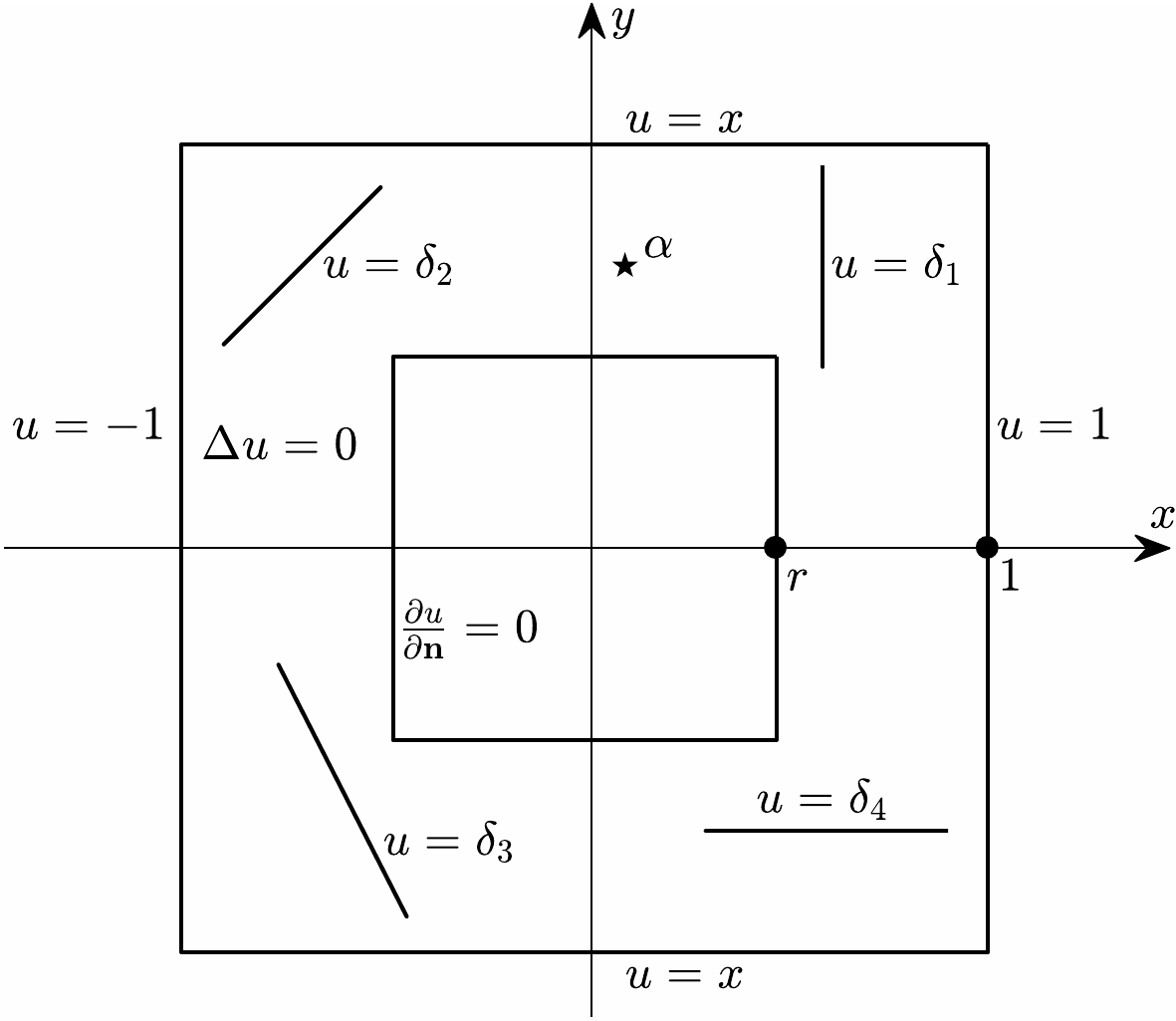}}
	}
	\caption{Geometry of the problem (for $m=4$).}
	\label{fig:dom-Dt}
\end{figure}

Our aim in this paper is to study the steady heat conduction problem within a thermal isotropic and homogeneous square medium that contains an inner isolation square. The body of the medium is a composite reinforced by non-overlapping CNTs having random orientation and position.  The CNTs are modeled as rigid line inclusions which are denoted by rectilinear slits $L_1,\ldots,L_m$. 
By denoting the inner square by $L_{m+1}$ and the outer square by $L_{m+2}$, the domain $D=(-1,1)\times(-1,1)\setminus(-r,r), \; 0<r<1$ occupied by the medium is multiply connected of connectivity $m+2$ and bordered by $L=\bigcup_{k=1}^{m+2}L_k$ (see Figure~\ref{fig:dom-Dt}).
We suppose that the temperature distribution $U(x,y)$ is known on the outer boundary $L_{m+2}$.  Furthermore, due to the remarkable high heat conductivity of CNTs, we treat them as heat superconductors and assume their temperature distribution to be  fixed to an indeterminate constant value along each line $L_k$ for $k=1\ldots,m$. We suppose also there is no net flux through each CNT.
Finally, we assume the inner boundary $L_{m+1}$ is perfectly isolated, which means there is no heat flux through the inner square. The governing equations for the temperature distribution are specifically given by the following mixed Dirichlet-Neumann boundary value problem: 
\begin{subequations}\label{eq:mix-bd-U}
	\begin{align}
	\label{eq:bvp-1-U}
	\Delta U &= 0, \quad \mbox{in }D, \\
	\label{eq:bvp-2-U}
	U &= \delta_k, \quad \mbox{on }L_k, \quad k=1,2,\ldots,m, \\
	\label{eq:bvp-3-U}
	\int_{L_k}\frac{\partial U}{\partial\bn}ds &= 0, \quad \quad k=1,2,\ldots,m, \\
	\label{eq:bvp-4-U}
	\frac{\partial U}{\partial\bn}&= 0, \quad \mbox{on }L_{m+1},\\
	\label{eq:bvp-5-U}
	U&= x, \quad \mbox{on }L_{m+2},
	\end{align}
\end{subequations}
where $\partial U/\partial\bn$ denotes the normal derivative of $U$, and $\delta_1,\ldots,\delta_m$ are undetermined real constants that need to be found alongside the real function $U$ (see Figure~\ref{fig:dom-Dt}).

Letting $z=x+\i y\in D$, we shall write for simplicity $U(z)$ instead of $U(x,y)$.
Note that the condition~\eqref{eq:bvp-5-U} means $U=-1$ on the left-side of the outer square and $U=1$ on its right-side. For the upper and lower sides, the temperature $U$ increases linearly and continuously from $-1$ to $+1$ as $x$ increases from $-1$ to $+1$. Moreover, we can write the boundary condition~\eqref{eq:bvp-5-U}  as $U(\xi)=\Re[\xi]$ for $\xi\in L_{m+2}$.

As the temperature distribution is unknown on the interface of CNTs (undetermined constants $\delta_k, \; k=1,\ldots,m$), the total degrees of freedom increase with the number of CNTs. An efficient approach to solve such large scale problem is to apply the numerical method presented in~\cite{Nas-ETNA} which is based on the boundary integral equations with the generalized Neumann kernel. However, since the boundary components $L_1,\ldots,L_m$ are slits, the above mixed boundary value problem is not directly solvable by this method. Using the invariance of Problem \eqref{eq:mix-bd-U} under conformal mapping, we need first to map the physical domain $D$ in the $z$-plane onto a new domain $G$ in the $w$-plane bordered by piecewise smooth Jordan curves, and then apply the integral equation method in the new domain $G$. In order to find the domain $G$, we employ the iterative method given in~\cite{Nas-Gre} as explained in the next section.

\section{Conformally mapping the physical domain}
\label{sc:dom}

\begin{figure}[ht] %
	\centerline{
		\scalebox{0.5}{\includegraphics[trim=0 0 0 0,clip]{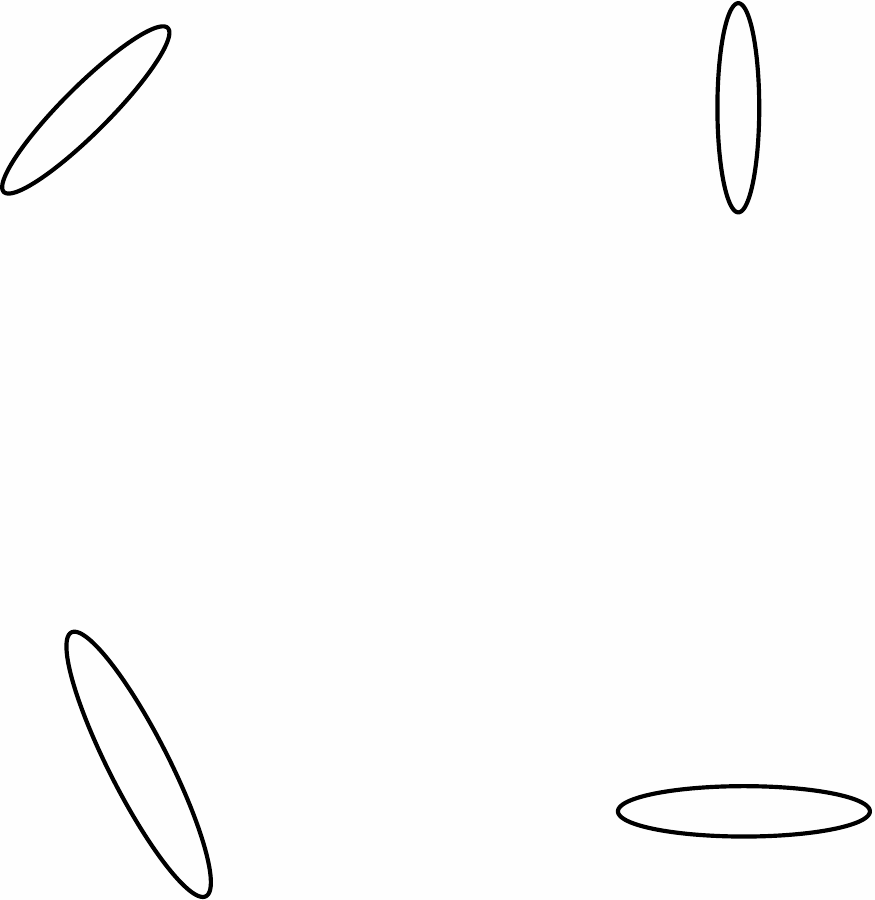}}
		\hfill
		\scalebox{0.5}{\includegraphics[trim=0 0 0 0,clip]{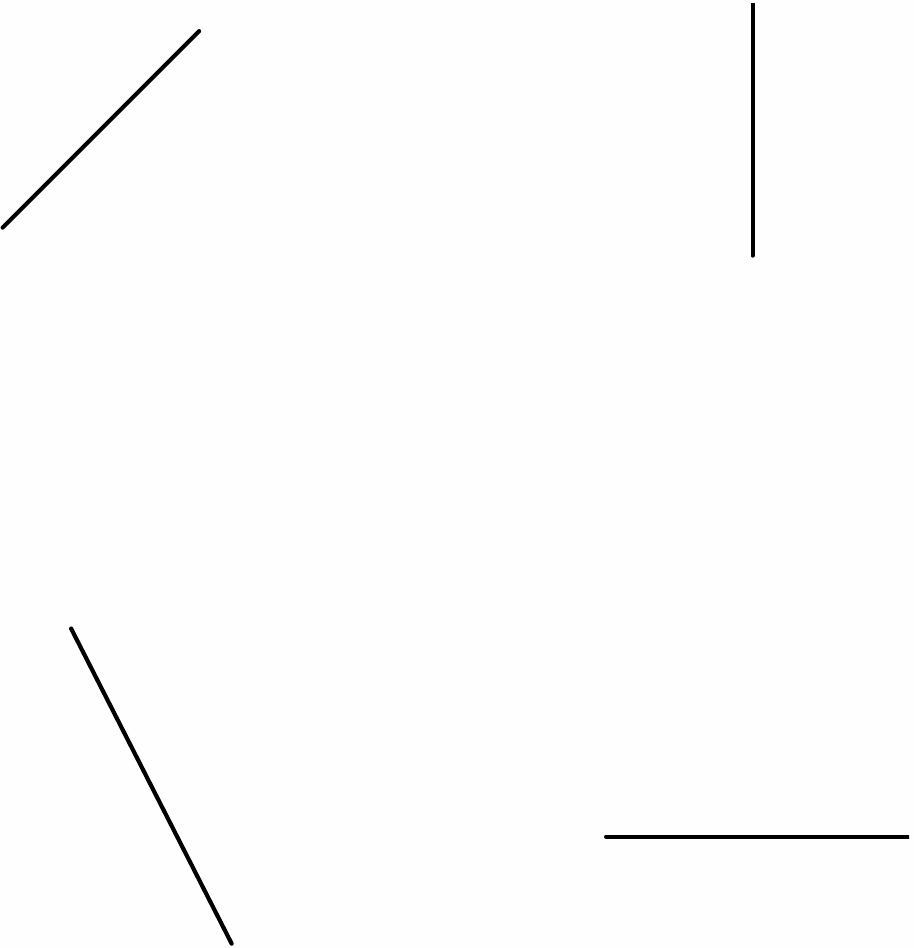}}
	}
	\caption{The domain $\hat D$ exterior to $m=4$ rectilinear slits (right) and the preimage domain $\hat G$ exterior to $4$ ellipses (left).}
	\label{fig:dom-D-map}
\end{figure}

Consider the unbounded multiply connected domain $\hat D$ being exterior to the $m$ rectilinear slits $L_1,\ldots,L_m$ (see Figure~\ref{fig:dom-D-map} (right)). We assume each slit $L_k$ make an angle $\beta_k$ with the positive $x$-axis for $k=1,2,\ldots,m$. The iterative method presented in~\cite{Nas-Gre} computes an unbounded preimage domain $\hat G$ in the exterior of $m$ ellipses $\Gamma_1,\ldots,\Gamma_m$ (see Figure~\ref{fig:dom-D-map} (left)). 
The ellipses are parameterized by
\begin{equation}\label{eq:eta-k}
\eta_k(t)=z_k+0.5e^{\i\beta_k}a_k(\cos t-\i r\sin t), \quad t\in J_k, \quad k=1,2,\ldots,m,
\end{equation}
where $0<r\le1$ is the ratio between minor and major axes' lengths of the ellipses. The parameters $z_k$ and $a_k$, $k=1,2,\ldots,m$, are computed by the iterative method. The method provides also the boundary values of the unique conformal mapping $z=\Phi(w)$ from the domain $\hat G$ in $w$-plane onto the domain $\hat D$ in $z$-plane with the normalization 
\begin{equation}\label{eq:map-norm}
\Phi(\infty)=\infty, \quad \lim_{w\to\infty}(\Phi(w)-w)=0.
\end{equation}
For more details, we refer the reader to~\cite{Nas-Gre,Nas-jsc19}

Putting $\hat\Gamma:=\partial \hat G = 
\hat\Gamma=\Gamma_1\cup\cdots\cup\Gamma_m$, the domain $\hat G$ is on the left of $\hat\Gamma$ because the curves $\Gamma_1,\ldots,\Gamma_m$ are clockwise oriented in view of \eqref{eq:eta-k}. 
According to~\eqref{eq:map-norm} the mapping function $
\hat\Phi(w)=\Phi(w)-w$
is analytic in $\hat G$ with $\hat\Phi(\infty)=0$.
Thus, the Cauchy integral formula implies that the values of the mapping function $z=\Phi(w)$ can be obtained for $w\in\hat G$ by 
\begin{equation*}\label{eq:cau-w1}
z=\Phi(w)=w+\frac{1}{2\pi\i}\int_{\hat\Gamma}\frac{\Phi(\tau)-\tau}{\tau-w}d\tau.
\end{equation*}
We can also use the Cauchy integral formula to determine the values of the inverse mapping function $w=\Phi^{-1}(z)$ for $z\in\hat D$ (see e.g.,~\cite{Nas-proc}). The inverse mapping function $\Phi^{-1}$ satisfies also the normalization~\cite[p.~114, p.~127]{Wen92} 
\begin{equation*}\label{eq:map-norm-i}
\Phi^{-1}(\infty)=\infty, \quad \lim_{z\to\infty}(\Phi^{-1}(z)-z)=0.
\end{equation*}
Therefore the values of $w=\Phi^{-1}(z)$ for $z\in\hat D$ can be computed by
\begin{equation}\label{eq:cau-z1}
w=\Phi^{-1}(z)=z+\frac{1}{2\pi\i}\int_{\hat L}\frac{\Phi^{-1}(\xi)-\xi}{\xi-z}d\xi.
\end{equation}

For the parameterization of the boundary $\hat\Gamma$, we adopt the same notations used in~\cite{Nas-log,Nas-jmaa11,Nas-jsc19,Nas-ETNA,Nas-Gre,Weg-Nas}. Let 
\[
\hat J = \bigsqcup_{k=1}^{m} J_k=\bigcup_{k=1}^{m}\{(t,k)\;:\;t\in J_k\},
\]
where $J_k=[0,2\pi]$ and $(t,k)\in\hat J$ means $t\in J_k$ for $k=1,2,\ldots,m$. In this way, we define a parameterization of the boundary $\hat\Gamma$ on $\hat J$ by
\begin{equation*}\label{e:eta-1}
\hat\eta(t,k)=\eta_k(t), \quad t\in J_k,\quad k=1,\ldots,m.
\end{equation*}
We will drop  the index $k$ in the notation of $\hat\eta$ as the value of $k$ such that $t\in J_k$ will be clear from the context. Therefore we simply write
\[
\hat\eta(t)= \left\{ \begin{array}{l@{\hspace{0.5cm}}l}
\eta_1(t),&t\in J_1,\\
\hspace{0.3cm}\vdots\\
\eta_m(t),&t\in J_m.
\end{array}
\right.
\]

The iterative method in~\cite{Nas-Gre} provides us with the boundary values $\Phi(\hat\eta(t))$, which are used to get a parameterization of the boundary $\hat L$ of $\hat D$ 
\begin{equation*}\label{eq:zet-1}
\hat\zeta(t)=\Phi(\hat\eta(t)), \quad t\in \hat J.
\end{equation*}
Thus, by making use of~\eqref{eq:cau-z1}, we can determine the values of the inverse map $z=\Phi^{-1}(w)$ for $w\in\hat D$ through
\begin{align}\label{eq:cau-z2}
w=\Phi^{-1}(z)
&=z+\frac{1}{2\pi\i}\int_{\hat J}\frac{\Phi^{-1}(\hat\zeta(t))-\hat\zeta(t)}{\hat\zeta(t)-z}\hat\zeta'(t)dt\nonumber\\
&=z+\frac{1}{2\pi\i}\int_{\hat J}\frac{\hat\eta(t)-\hat\zeta(t)}{\hat\zeta(t)-z}\hat\zeta'(t)dt
\end{align}
To compute the values of $\hat\zeta'(t)$, we first approximate the real and imaginary parts of $\hat\zeta(t)$ on each interval $J_k$, $k=1,2,\ldots,m$, by trigonometric interpolating polynomials, and then differentiate.

\begin{figure}[ht] %
	\centerline{
		\scalebox{0.45}{\includegraphics[trim=0 0 0 0,clip]{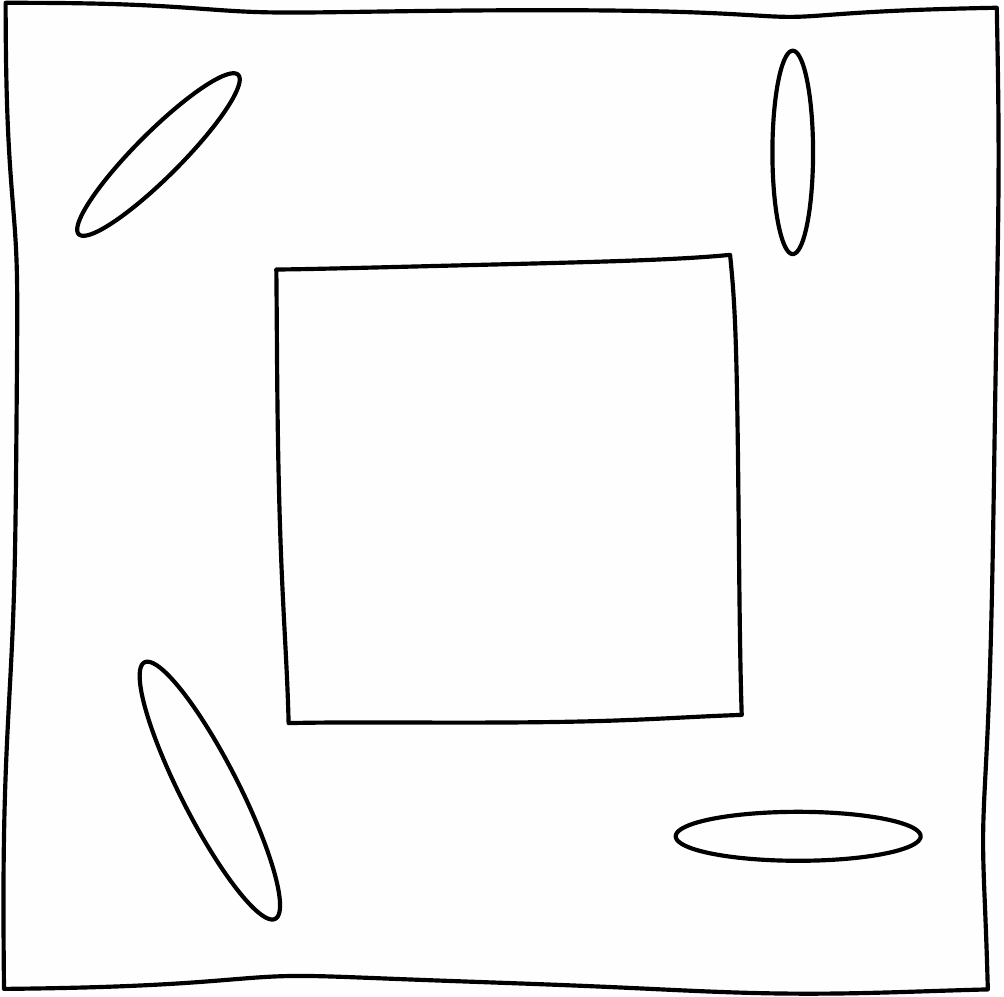}}
		\hfill
		\scalebox{0.45}{\includegraphics[trim=0 0 0 0,clip]{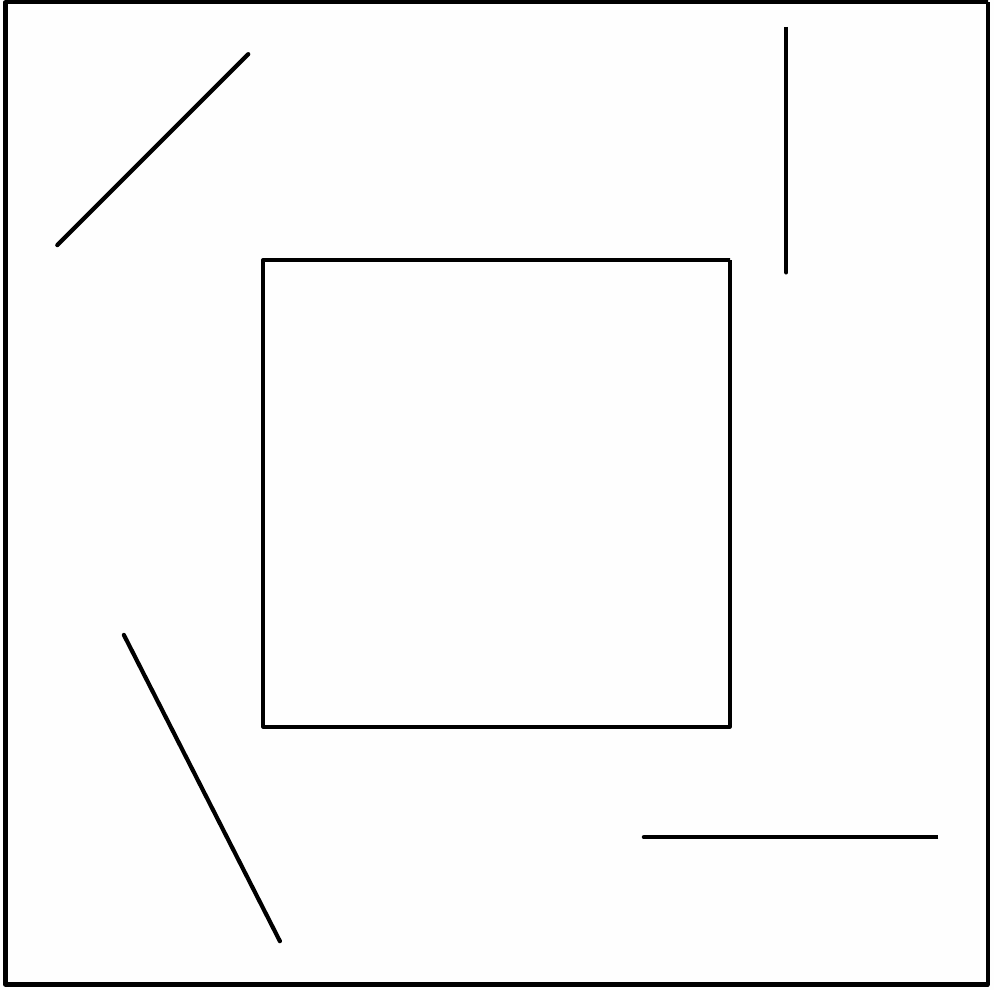}}
	}
	\caption{The given domain $D$ (right) and the preimage domain $G$ (left) for $m=4$.}
	\label{fig:dom-pre}
\end{figure}

Now we are ready to compute the preimage of the given domain $D$ (see Figure~\ref{fig:dom-pre} (right)). The inverse mapping function $w=\Phi^{-1}(z)$ maps the domain outside the slits $L_1,\ldots,L_m$ onto the domain outside of the ellipses $\Gamma_1,\ldots,\Gamma_m$. Furthermore, $\Phi^{-1}$ maps the squares $L_{m+1}$ and $L_{m+2}$ onto piecewise smooth Jordan curves $\Gamma_{m+1}$ and $\Gamma_{m+2}$, respectively, such that the $m$ ellipses are in the ring domain between the curves $\Gamma_{m+1}$ and $\Gamma_{m+2}$ (see Figure~\ref{fig:dom-pre} (left)). Thus the whole boundary of the domain $G$ is $
\Gamma=\partial G=\Gamma_1\cup\cdots\cup\Gamma_{m+2}$.

For each $k=m+1,m+2$, we parameterize the square $L_{k}$ by $\zeta_k(t)$, $t\in J_k=[0,2\pi]$, where $\zeta_k(t)$ is chosen using the same approach used in~\cite[pp.~696-697]{Nas-log}. Hence for $k=m+1,m+2$, the curve $\Gamma_k$ is parameterized by
\[
\eta_k(t)=\Phi^{-1}(\zeta_k(t)), \quad t\in J_k,
\]
such that the values of $\Phi^{-1}(\zeta_k(t))$ are computed through~\eqref{eq:cau-z2}. The values of the derivative $\eta'_k(t)$, for $k=m+1,m+2$, are computed numerically again by using trigonometric interpolating polynomials as explained above. Recall that the parameterization of the ellipses are given by~\eqref{eq:eta-k}. Henceforth if we take $J$ as the disjoint union of the $m+2$ intervals $J_1,\ldots,J_{m+2}$, then we parameterize the whole boundary $\Gamma$ by
\begin{equation*}\label{eq:eta}
\eta(t)= \left\{ \begin{array}{l@{\hspace{0.5cm}}l}
\eta_1(t),&t\in J_1,\\
\hspace{0.3cm}\vdots\\
\eta_{m+2}(t),&t\in J_{m+2}.
\end{array}
\right.
\end{equation*}

\section{Computing the temperature distribution}
\label{sc:U}

Note first that the boundary value problem~\eqref{eq:mix-bd-U} is invariant under conformal mapping. Indeed, the unique solution $U$ of the problem~\eqref{eq:mix-bd-U} is given by
\begin{equation}\label{eq:U-u}
U(z)=u(\Phi^{-1}(z)), \quad z\in D,
\end{equation}
such that $u$ is the unique solution of the following boundary value problem on the domain $G$ in $w$-plane,
\begin{subequations}\label{eq:mix-bd}
	\begin{align}
	\label{eq:bvp-1}
	\Delta u &= 0, \quad \mbox{in }G, \\
	\label{eq:bvp-2}
	u &= \delta_k, \quad \mbox{on }\Gamma_k, \quad k=1,2,\ldots,m, \\
	\label{eq:bvp-3}
	\int_{\Gamma_k}\frac{\partial u}{\partial\bn}ds &= 0, \quad \quad k=1,2,\ldots,m, \\
	\label{eq:bvp-4}
	\frac{\partial u}{\partial\bn}&= 0, \quad \mbox{on }\Gamma_{m+1},\\
	\label{eq:bvp-5}
	u(\tau)&= \Re[\tau] \quad \mbox{for}\;\tau\in\Gamma_{m+2}.
	\end{align}
\end{subequations}
As $u$ is harmonic, we can write 
\begin{equation}\label{eq:u-f}
u(w)=\Re[f(w)]
\end{equation}
where $f(w)$ is a single-valued analytic function on the domain $G$. Assuming the boundary values of the function $f$ takes the form
\begin{equation*}\label{eq:F-bdv}
f(\eta(t))=\phi(t)+\i\psi(t), \quad t\in J
\end{equation*}
on the boundary of $G$, it follows from~\eqref{eq:bvp-2} that 
\begin{equation}\label{eq:phi-k}
\phi(t)=\delta_k, \quad t\in J_k, \quad k=1,2,\ldots,m.
\end{equation}
Similarly, equation~\eqref{eq:bvp-5} implies that the values of the function $\phi(t)$ are known for $t\in J_{m+2}$ with
\begin{equation}\label{eq:phi-m+2}
\phi(t)=\Re[\eta_{m+2}(t)], \quad t\in J_{m+2}.
\end{equation}
No information can be found from~\eqref{eq:mix-bd} about the values of the function $\phi(t)$ for $t\in J_{m+1}$. Nevertheless, the condition~\eqref{eq:bvp-4} yields $\psi'(t)=0$ for $t\in J_{m+1}$~\cite{Mit-chap,Nas-amc}. Thus, $\psi(t)$ is a constant function on the interval $J_{m+1}$, i.e., 
\begin{equation}\label{eq:psi-m+1}
\psi(t)=\delta_{m+1}, \quad t\in J_{m+1}.
\end{equation}
where $\delta_{m+1}$ is an undetermined real constant.

Let us define the function $\hat A(t)$ on $J$ by
\begin{equation*}\label{eq:hA}
\hat A(t)=\left\{ \begin{array}{l@{\hspace{0.5cm}}l}
1,     &t\in J_1,\\
\hspace{0.3cm}\vdots\\
1,     &t\in J_{m}, \\
-\i,   &t\in J_{m+1},\\
1,     &t\in J_{m+2}.
\end{array}
\right.
\end{equation*}
Following~\eqref{eq:phi-k}, \eqref{eq:phi-m+2} and~\eqref{eq:psi-m+1}, the function $F$ satisfies the boundary conditions
\begin{equation}\label{eq:F-bdc}
\Re[\hat A(t)f(\eta(t))]=\left\{ \begin{array}{l@{\hspace{0.5cm}}l}
\delta_1,     &t\in J_1,\\
\hspace{0.3cm}\vdots\\
\delta_m,     &t\in J_{m}, \\
\delta_{m+1},   &t\in J_{m+1},\\
\Re[\eta_{m+2}(t)],     &t\in J_{m+2},
\end{array}
\right.
\end{equation}
which are known as the Riemann-Hilbert problem. To solve the problem~\eqref{eq:F-bdc}, we tale an auxiliary given point $\alpha$ in the domain $G$. Since we are only interested in determining the real function $u=\Re f$, we can assume that $f(\alpha)=c$ is real. We define an auxiliary function $g$ on the domain $G$ by
\begin{equation}\label{eq:g}
g(w)=\frac{f(w)-c}{w-\alpha}, \quad w\in G\cup\Gamma.
\end{equation}
The function is clearly analytic in $G$. We consider also a complex-valued function $A(t)$ defined on $J$ by
\begin{equation*}\label{eq:A}
A(t) = e^{-\i\theta(t)}(\eta(t)-\alpha), \quad t\in J,
\end{equation*}
where $\theta(t)$ is the piecewise constant function given by
\[
\theta(t)=\left\{ \begin{array}{l@{\hspace{0.5cm}}l}
0,     &t\in J_1,\\
\hspace{0.3cm}\vdots\\
0,     &t\in J_{m}, \\
\pi/2,   &t\in J_{m+1},\\
0,     &t\in J_{m+2}.
\end{array}
\right.
\]
Therefore the boundary conditions in~\eqref{eq:F-bdc} implies that $g(w)$ solves the Riemann-Hilbert problem
\begin{equation*}\label{eq:rhp}
\Re[A(t)g(\eta(t))]=\gamma(t)+h(t), \quad t\in J,
\end{equation*}
given that 
\[
\gamma(t)=\left\{ \begin{array}{l@{\hspace{0.5cm}}l}
0,     &t\in J_1,\\
\hspace{0.3cm}\vdots\\
0,   &t\in J_{m+1},\\
\Re[\eta_{m+2}(t)],     &t\in J_{m+2},
\end{array}
\right. \quad
h(t)=\left\{ \begin{array}{l@{\hspace{0.5cm}}l}
h_1,     &t\in J_1,\\
h_2,     &t\in J_2,\\
\hspace{0.3cm}\vdots\\
h_{m+2},     &t\in J_{m+2},
\end{array}
\right.
\]
where $h_k=\delta_k-c$ for $k=1,\ldots,m$, $h_{m+1}=\delta_{m+1}$, and $h_{m+2}=-c$. The piecewise constant function $h$ is an unknown and need to be determined together with the function $f$. Both of these two functions can be found using a boundary integral equation method based on the generalized Neumann kernel. More precisely, following~\cite{Nas-jmaa11} the boundary values of the analytic function $g$ are determined through
\begin{equation*}\label{eq:Af}
A(t)g(\eta(t))=\gamma(t)+h(t)+\i\mu(t), \quad  t\in J,
\end{equation*}
where $\mu(t)$ uniquely solves the integral equation
\begin{equation}\label{eq:ie}
(\bI-\bN)\mu=-\bM\gamma
\end{equation}
and the function $h$ is computed by
\begin{equation}\label{eq:h}
h=[\bM\mu-(\bI-\bN)\gamma]/2.
\end{equation}
Here, $\bI$ is the identity operator and the integral operators $\bN$ and $\bM$ are respectively defined by
\begin{align*}
\bN\mu(s) = \int_J \frac{1}{\pi}\Im\left(
\frac{A(s)}{A(t)}\frac{\eta'(t)}{\eta(t)-\eta(s)}\right) \mu(t) dt, \quad s\in J,\\
\bM\mu(s) = \int_J \frac{1}{\pi}\Re\left(
\frac{A(s)}{A(t)}\frac{\eta'(t)}{\eta(t)-\eta(s)}\right) \mu(t) dt, \quad s\in J.
\end{align*}
For more details, see~\cite{Nas-jmaa11,Nas-ETNA}.

We compute approximations to the functions $\mu$ in~\eqref{eq:ie} and $h$ in~\eqref{eq:h} by the MATLAB function \verb|fbie| from~\cite{Nas-ETNA} as follows
\begin{verbatim}
[mu,h] = fbie(et,etp,A,gamk,n,iprec,restart,tol,maxit).
\end{verbatim}
This functions employs a discretization of the integral equation~(\ref{eq:ie}) by the Nystr\"om method using the trapezoidal rule to obtain an algebraic linear system~\cite{Atk97,Tre-Trap}. Each boundary component is discretized by $n$ nodes yielding a linear system of size $(m+2)n\times(m+2)n$. This system is solved by the MATLAB function $\mathtt{gmres}$ in which the matrix-vector multiplication is computed using the MATLAB function $\mathtt{zfmm2dpart}$ from $\mathtt{FMMLIB2D}$ toolbox~\cite{Gre-Gim12}. In our numerical calculations, we choose $\mathtt{iprec}=5$; which means the tolerance in the FMM is $0.5\times 10^{-15}$. The GMRES performs a maximum number of iterations {\tt maxit=100} with an accuracy tolerance {\tt tol=1e-14}, and the method is used without restart by choosing {\tt restart=[\,]}. 

By computing the real functions $\mu$ and $h$, we obtain the boundary values of the analytic function $g$ through~\eqref{eq:Af}. Then, in view of~\eqref{eq:g}, we obtain the boundary values of the analytic function $f$ by
\[
f(\eta(t))=(\eta(t)-\alpha)g(\eta(t))+c.
\]
The values of the function $f(\Phi^{-1}(z))$ can be computed for $z\in D$ by the Cauchy integral formula. Finally, following~\eqref{eq:U-u} and~\eqref{eq:u-f}, the values of the temperature distribution $U(z)$ is determined for $z\in D$ by
\begin{equation}\label{eq:U(z)}
U(z)=\Re\left[F(z))\right]
\end{equation}
where 
\begin{equation}\label{eq:F(z)}
F(z)=f(\Phi^{-1}(z)), \quad z\in D,
\end{equation}
is an analytic function on the medium $D$ called complex distribution temperature. The derivative if this function is usefully related to the heat flux as we will see in the next section.

\section{Computing the  heat flux}
\label{sc:q}

According to Fourier's law of conduction the heat flux vector $q$ related to the gradient temperature is given by the equation
$$
q=-\lambda\nabla U=-\lambda(\frac{\partial U}{\partial x},\frac{\partial U}{\partial y})
$$
where $\lambda $ is the equivalent thermal conductivity which we assume to be normalized to unity.

From~\eqref{eq:U(z)} and using the Cauchy-Riemann equations, it follows that the derivative of the complex temperature distribution $F(z)$ on $D$ is given by
\[
F'(z)=\frac{\partial U}{\partial x}-\i\frac{\partial U}{\partial y}. 
\]
One the other hand \eqref{eq:F(z)} yields
\[
F'(z)=\frac{f'(\Phi^{-1}(z))}{\Phi'(\Phi^{-1}(z))}, \quad z\in D,
\]
where the denominator does not vanish in the domain $D$ since $\Phi$ is a conformal mapping.  
Therefore the heat flux can be expressed in terms of $F'(z)$ by the formula
\begin{equation}\label{eq:q}
q(z)=-\left.\left(\frac{\partial U}{\partial x},\frac{\partial U}{\partial y}\right)\right|_z
=-\overline{F'(z)}
=-\overline{\left(\frac{f'(\Phi^{-1}(z))}{\Phi'(\Phi^{-1}(z))}\right)}, \quad z\in D.
\end{equation}

The values of the heat flux $q$ can be estimated on the domain $D$ by first computing the derivatives of the boundary values of the analytic functions $f$ and $\Phi$ on each boundary components using trigonometric interpolating polynomials as explained above. Then, the values of $f'(\Phi^{-1}(z))$ and $\Phi'(\Phi^{-1}(z))$, in the right-hand side of~\eqref{eq:q}, are computed for $z\in D$ using the Cauchy integral formula.

\section{Examples}
\label{sc:ex}

\begin{figure}[ht] %
	\centerline{
		\scalebox{0.45}{\includegraphics[trim=1cm 0 1cm 0,clip]{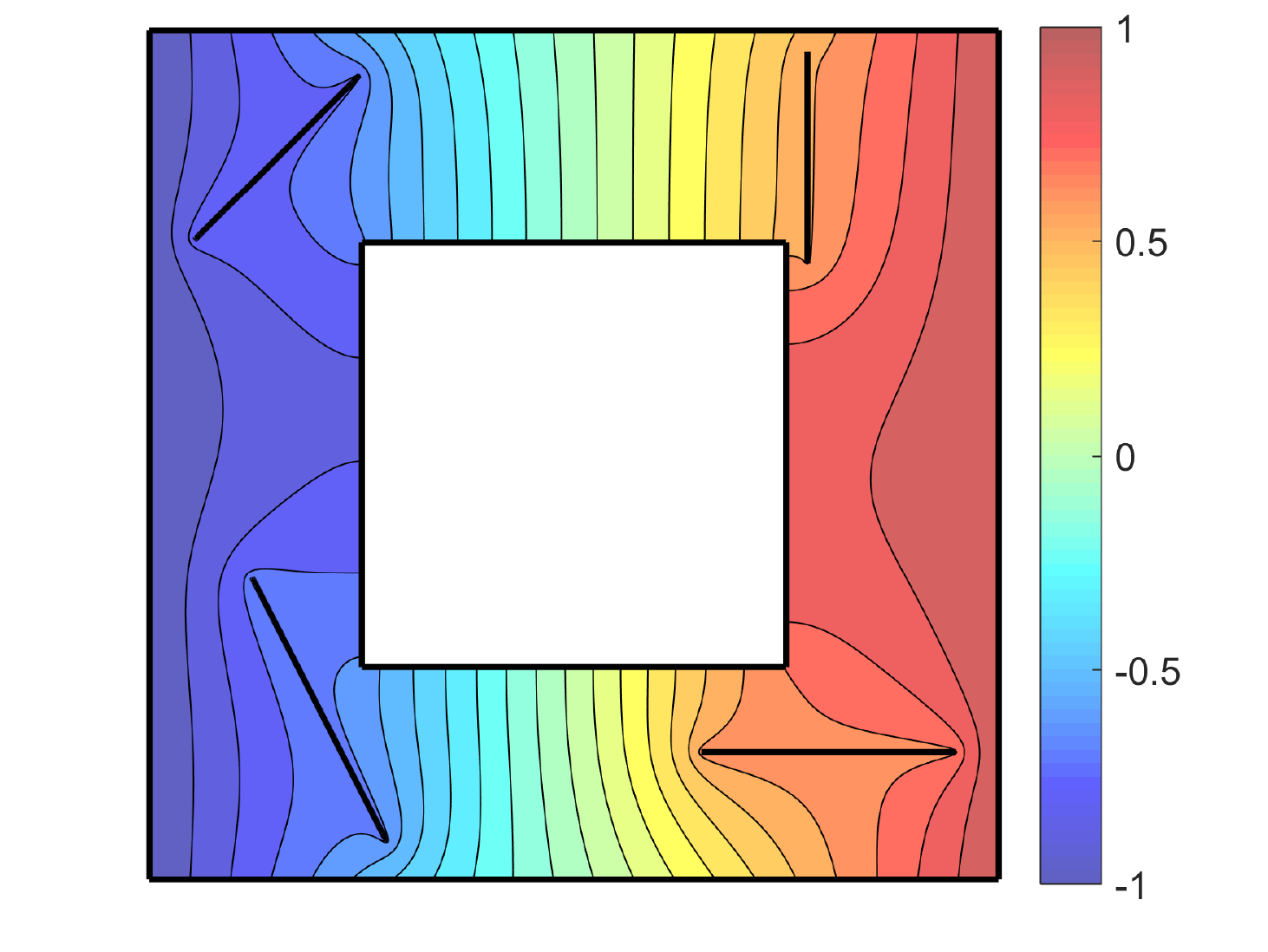}}
		\hfill
		\scalebox{0.45}{\includegraphics[trim=1cm 0 1cm 0,clip]{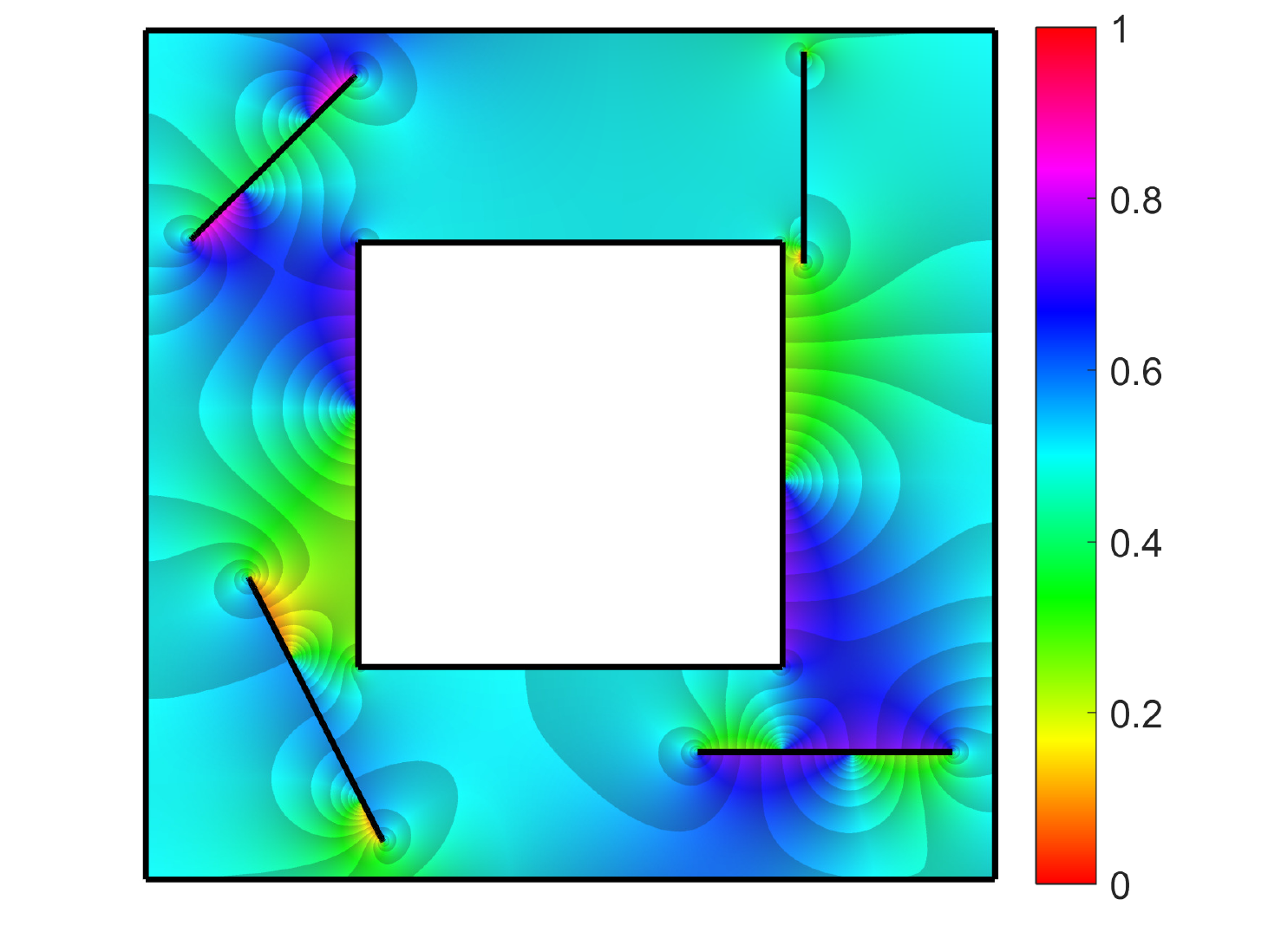}}
	}
	\caption{A contour plot of the temperature distribution $U$ (left) and a phase portrait of the heat flux $q$ (right) for $4$ CNTs and $r=0.5$.}
	\label{fig:4slits}
\end{figure}

\begin{figure}[ht] %
	\centerline{
		\scalebox{0.45}{\includegraphics[trim=1cm 0 1cm 0,clip]{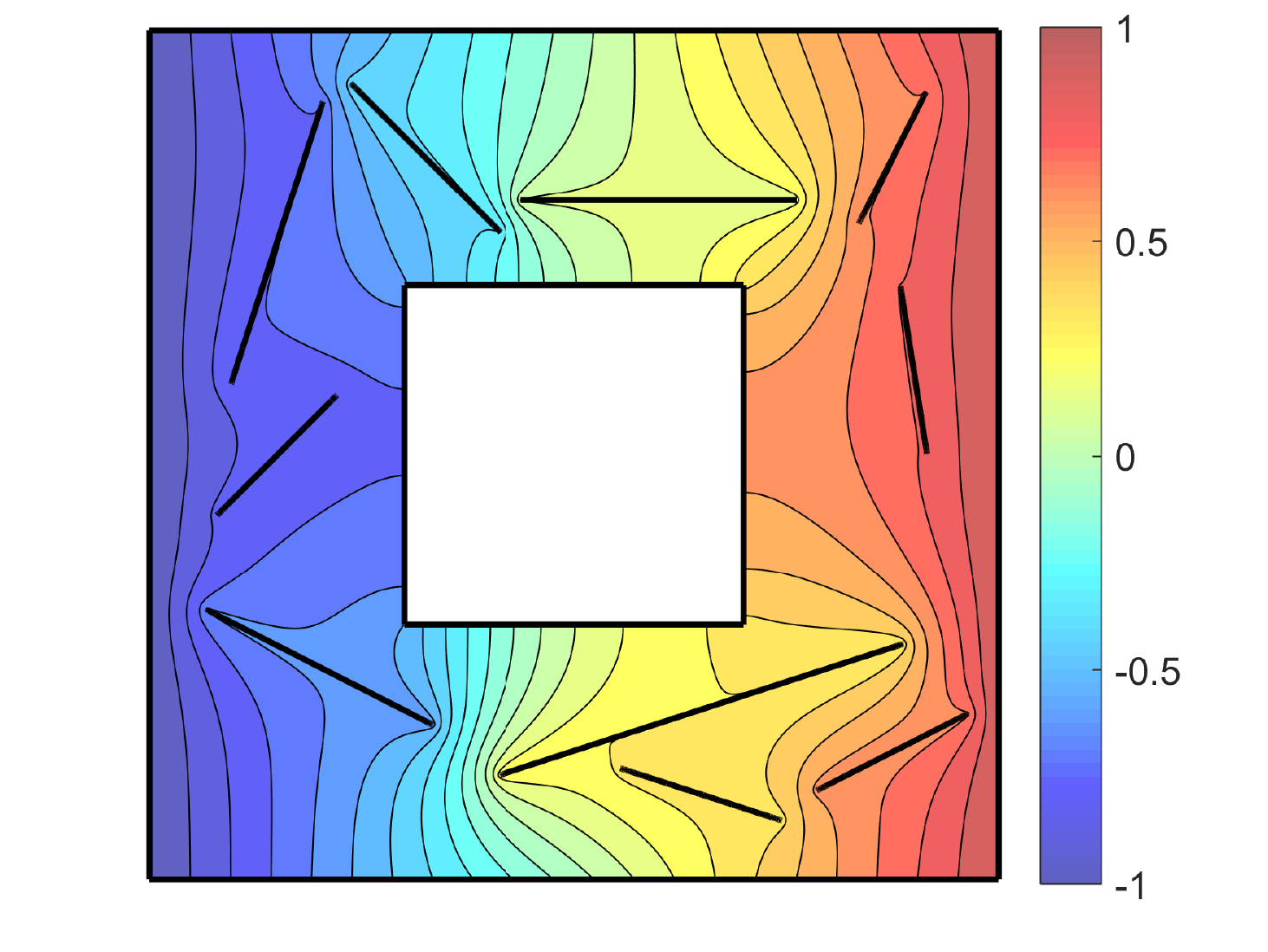}}
		\hfill
		\scalebox{0.45}{\includegraphics[trim=1cm 0 1cm 0,clip]{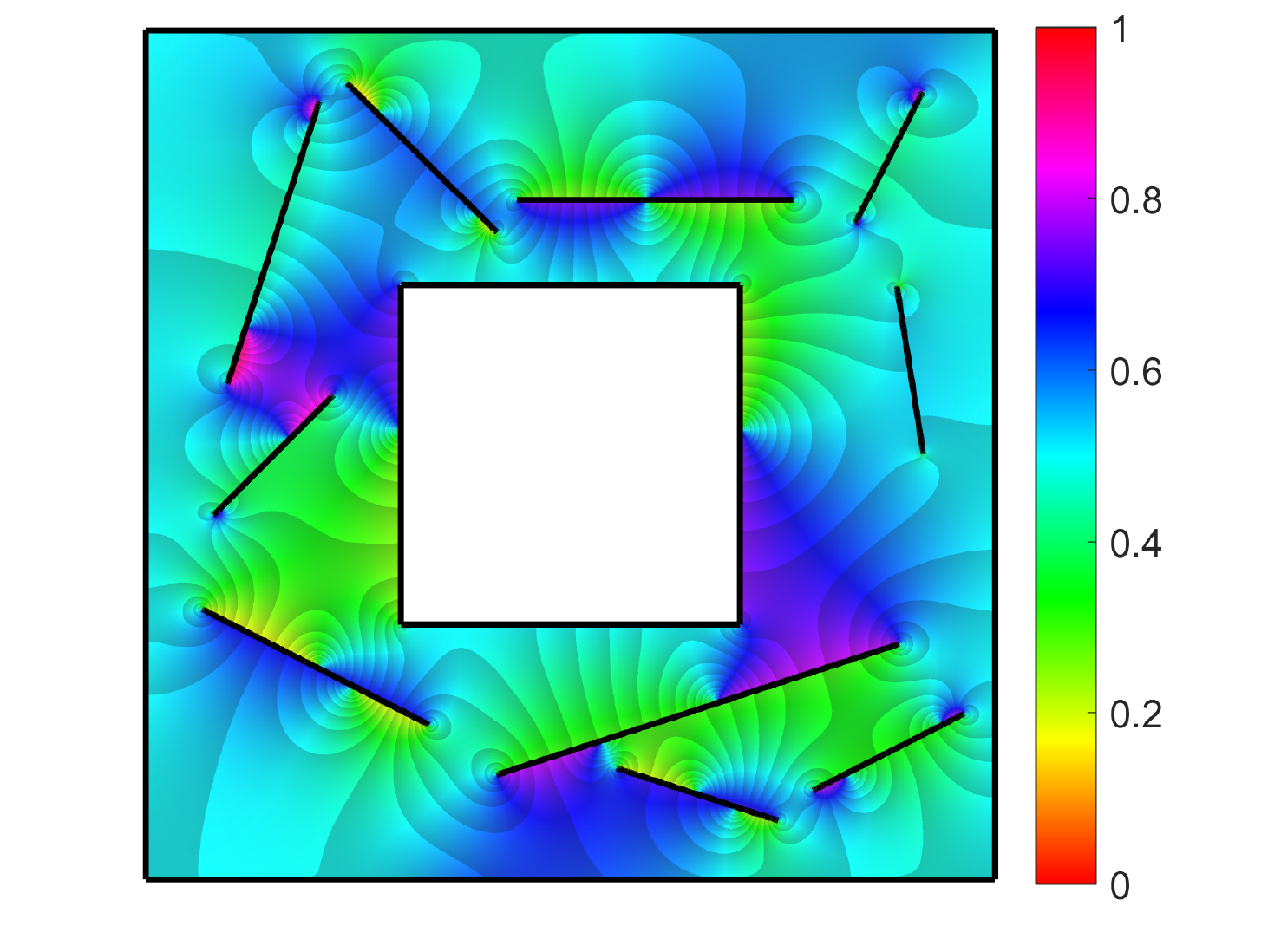}}
	}
	\caption{A contour plot of the temperature distribution $U$ (left) and a phase portrait of the heat flux $q$ (right) for $10$ CNTs and $r=0.5$.}
	\label{fig:10slits}
\end{figure}

\begin{figure}[ht] %
	\centerline{
		\scalebox{0.45}{\includegraphics[trim=1cm 0 1cm 0,clip]{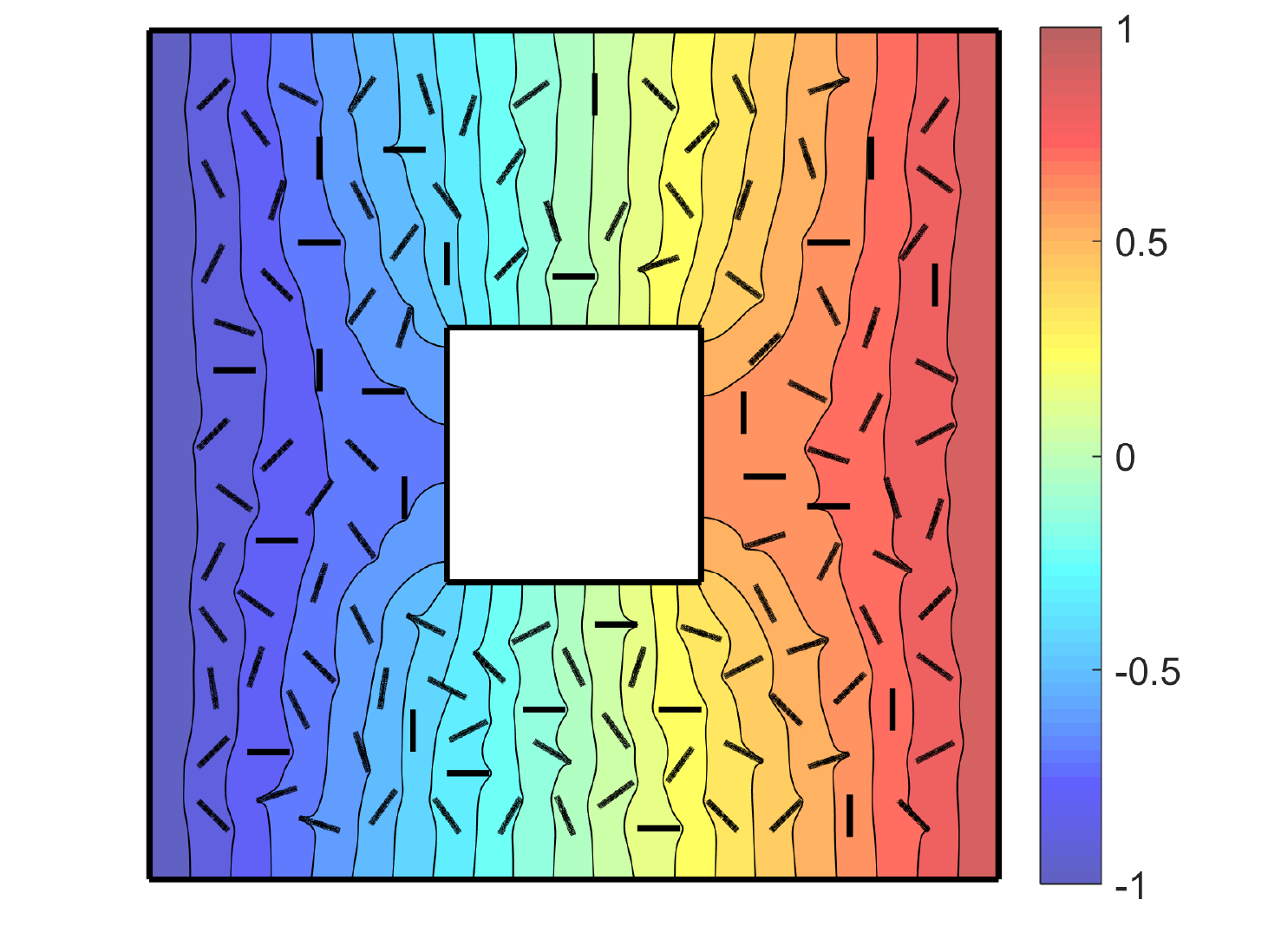}}
		\hfill
		\scalebox{0.45}{\includegraphics[trim=1cm 0 1cm 0,clip]{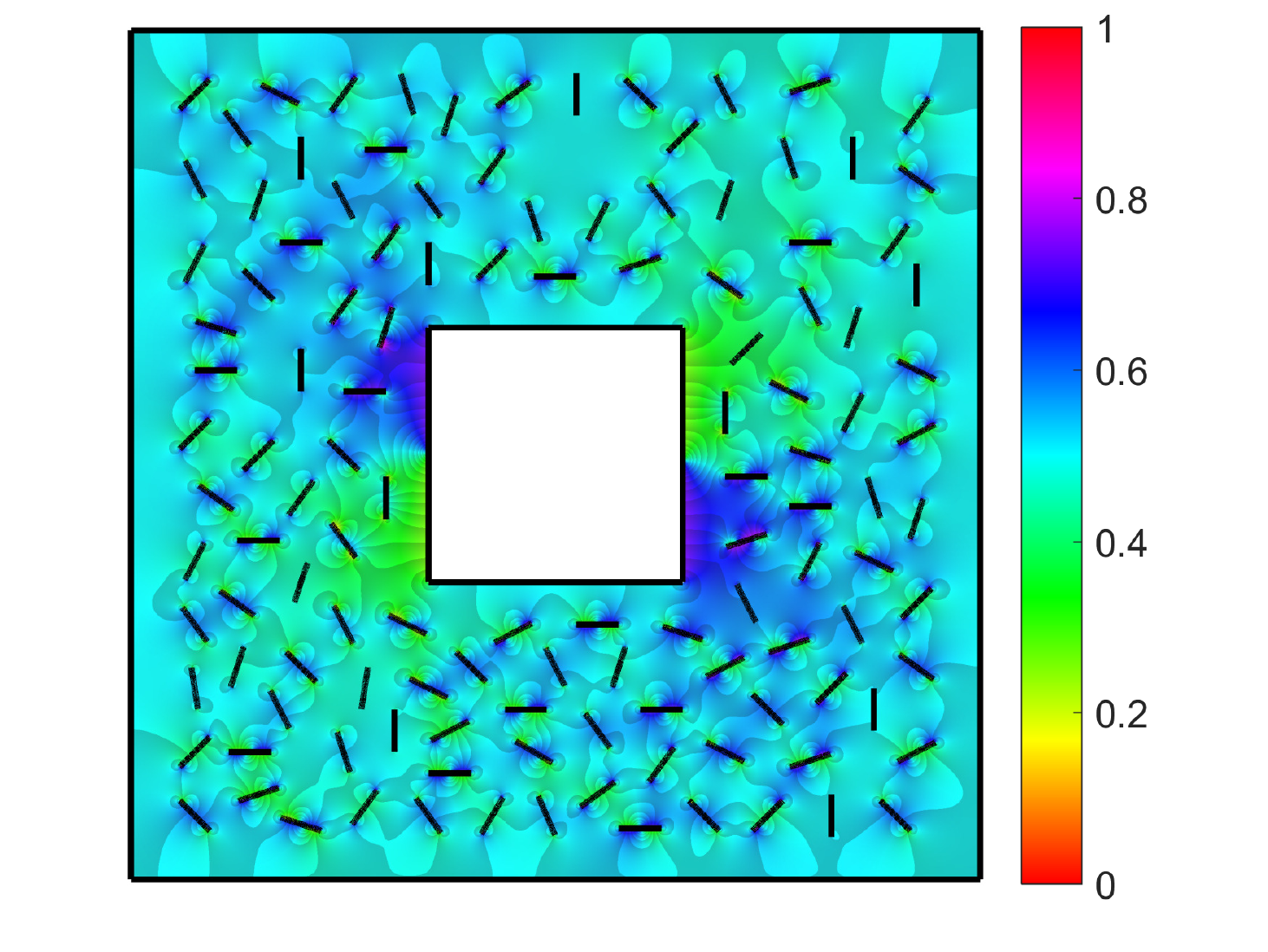}}
	}
	\caption{A contour plot of the temperature distribution $U$ (left) and a phase portrait of the heat flux $q$ (right) for $123$ CNTs with $r=0.3$.}
	\label{fig:123slits}
\end{figure}

We apply the method presented above to compute the temperature field $U$ and the heat flux $q$ for four different number of CNTs. 
In the first example, we consider the case of $4$ CNTs of different lengths with an inner square of side length $r=0.5$ (see Figure~\ref{fig:4slits}). 
In the second example, we have $10$ CNTs of different lengths and $r=0.4$ (see Figure~\ref{fig:10slits}).  
The third example involves $123$ CNTs of equal length $0.1$ with $r=0.3$  (see Figure~\ref{fig:123slits}).
In the fourth example, we consider $1005$ CNTs of random lengths between $0.02$ and $0.04$ with $r=0.1$  (see Figure~\ref{fig:123slits}). The centers and angles of the rigid line inclusions are chosen randomly so that all lines are non-overlapping.
For these fourth all examples, we use $n=2^{11}$.

Estimating the boundary values of the function $f$, requires two steps:
\begin{enumerate}
	\item Computing the domain $G$. For this step, the total number of discretization points is $mn$.
	\item Compute the boundary values of the function $f$. The total number of discretization points is $(m+2)n$.
\end{enumerate}
The total CPU time required by the two steps to compute $f(\eta(t))$ is shown in Table~\ref{tab:time}.

To compute the values of $U$ and $q$, we discretize the domain $G$ using a matrix \verb|W| with approximately $3$ million points in $G$. Thus ${\tt Z}=\Phi({\tt W})$ is a matrix of points that discretize the domain $D$. Afterwards, we compute the values of the temperature distribution $U$ and the heat flux $q$ at the points of the matrix ${\tt Z}$ using the formulas~\eqref{eq:U(z)} and~\eqref{eq:q}, respectively. For the temperature distribution $U$, we use the MATLAB function \verb|contourf| to plot the contours of the function $U$ on the domain $D$. To visualize the heat flux $q$, 
we present a phase portrait with modulus contour lines of $q$~\cite{Weg92}. The colors represent the phase of $q$ where red means $q$ is in the direction of the positive real axis, green for the positive imaginary axis, cyan for the negative real axis, and violet for the negative imaginary~\cite{Weg92}. We show also the contour lines of the amplitude of $q$.
The total CPU time required to compute the values of $U$ and $q$ are shown in Table~\ref{tab:time}.

\begin{figure}[ht] %
	\centerline{
		\scalebox{0.4}{\includegraphics[trim=0 0 0 0,clip]{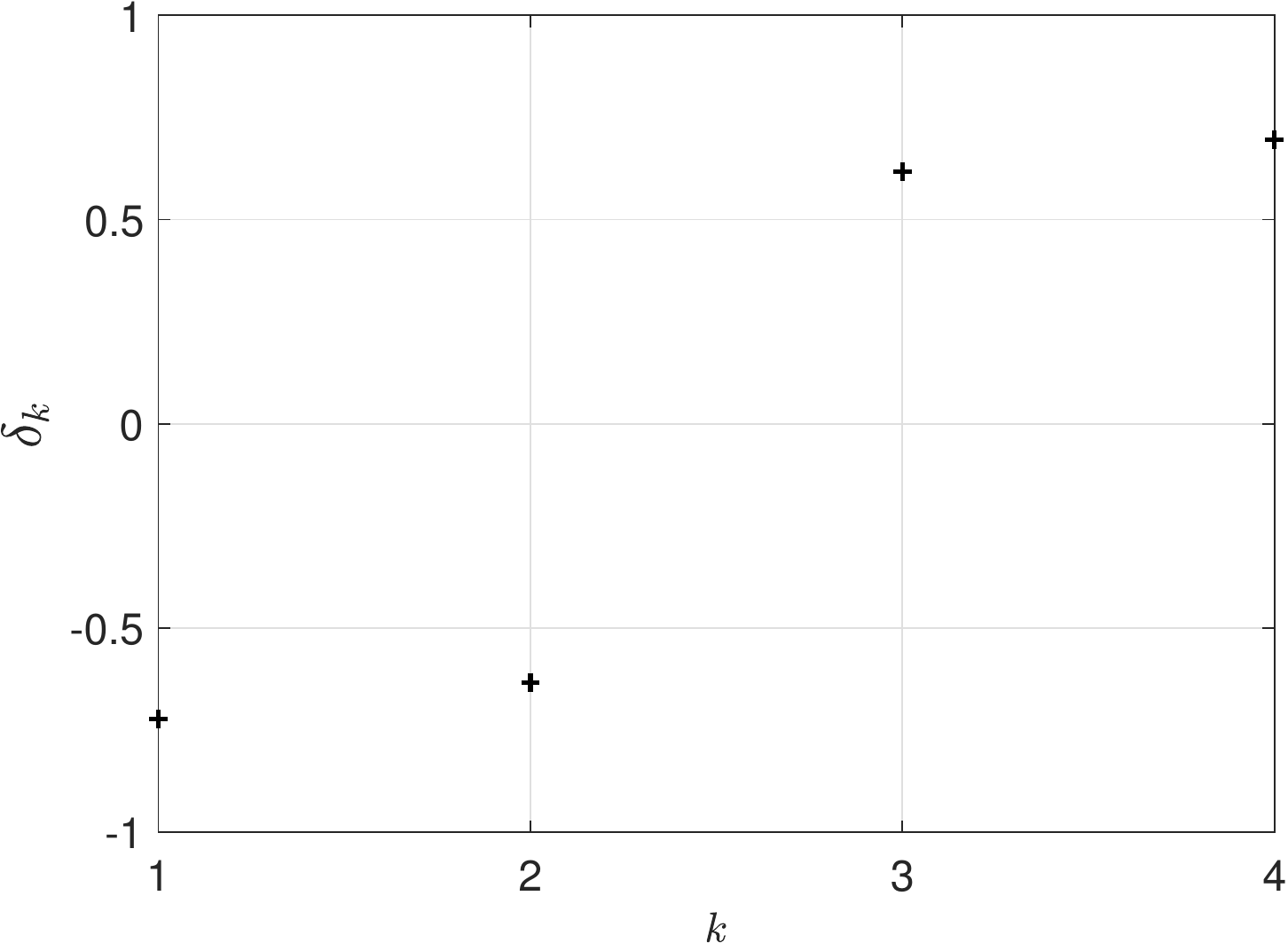}}
		\hfill
		\scalebox{0.4}{\includegraphics[trim=0 0 0 0,clip]{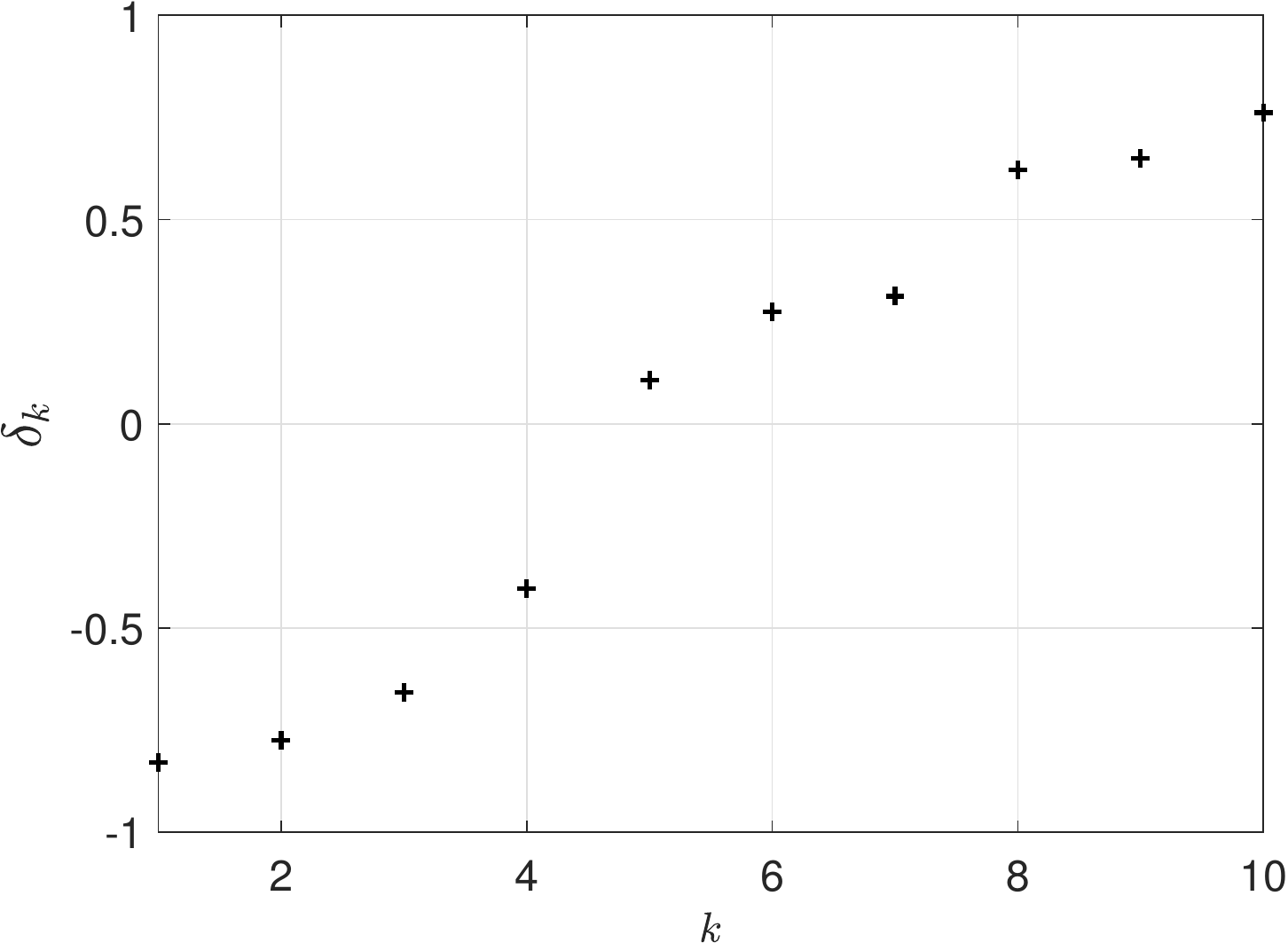}}
	}
	\centerline{
		\scalebox{0.4}{\includegraphics[trim=0 0 0 0,clip]{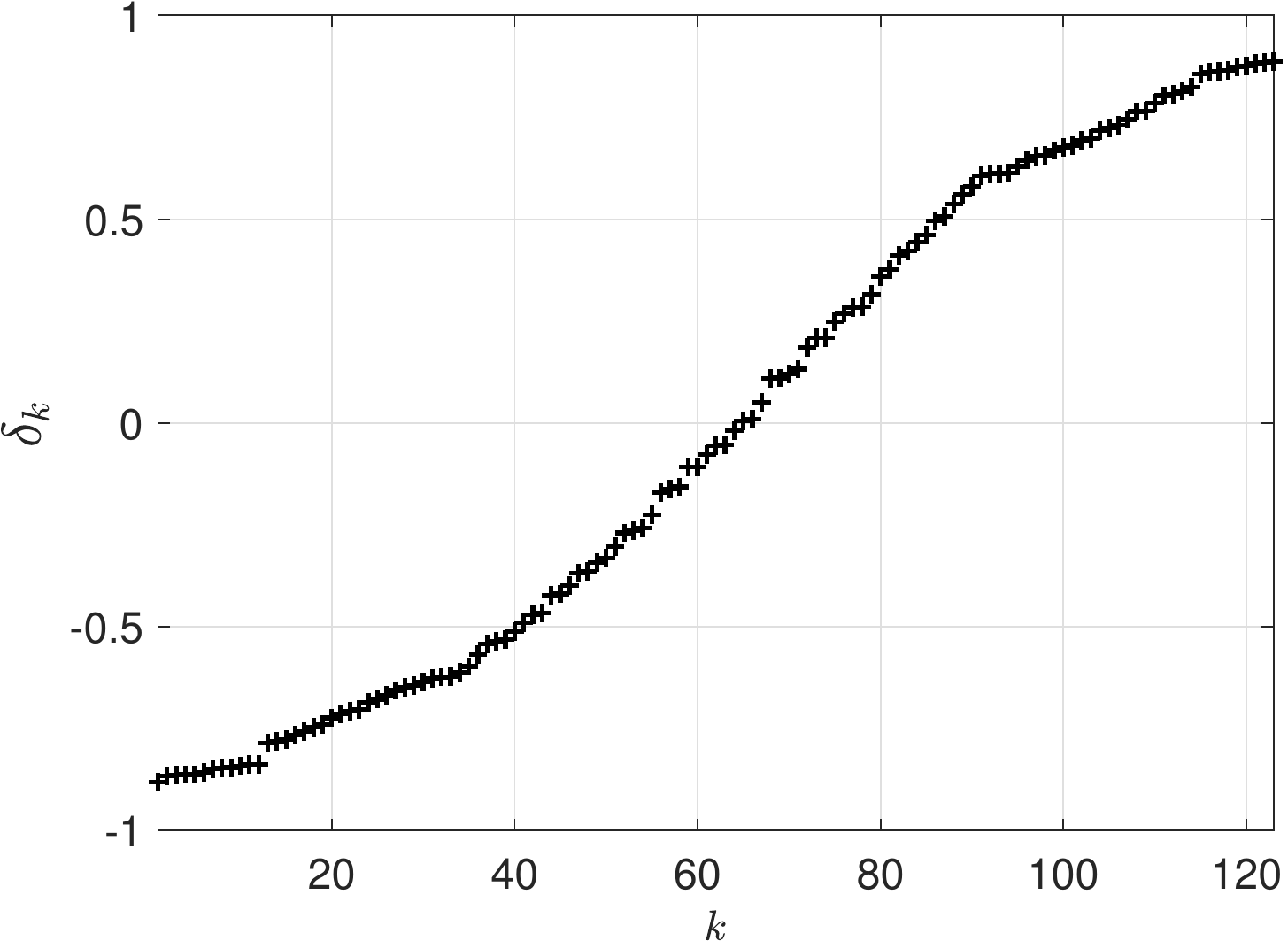}}
		\hfill
		\scalebox{0.4}{\includegraphics[trim=0 0 0 0,clip]{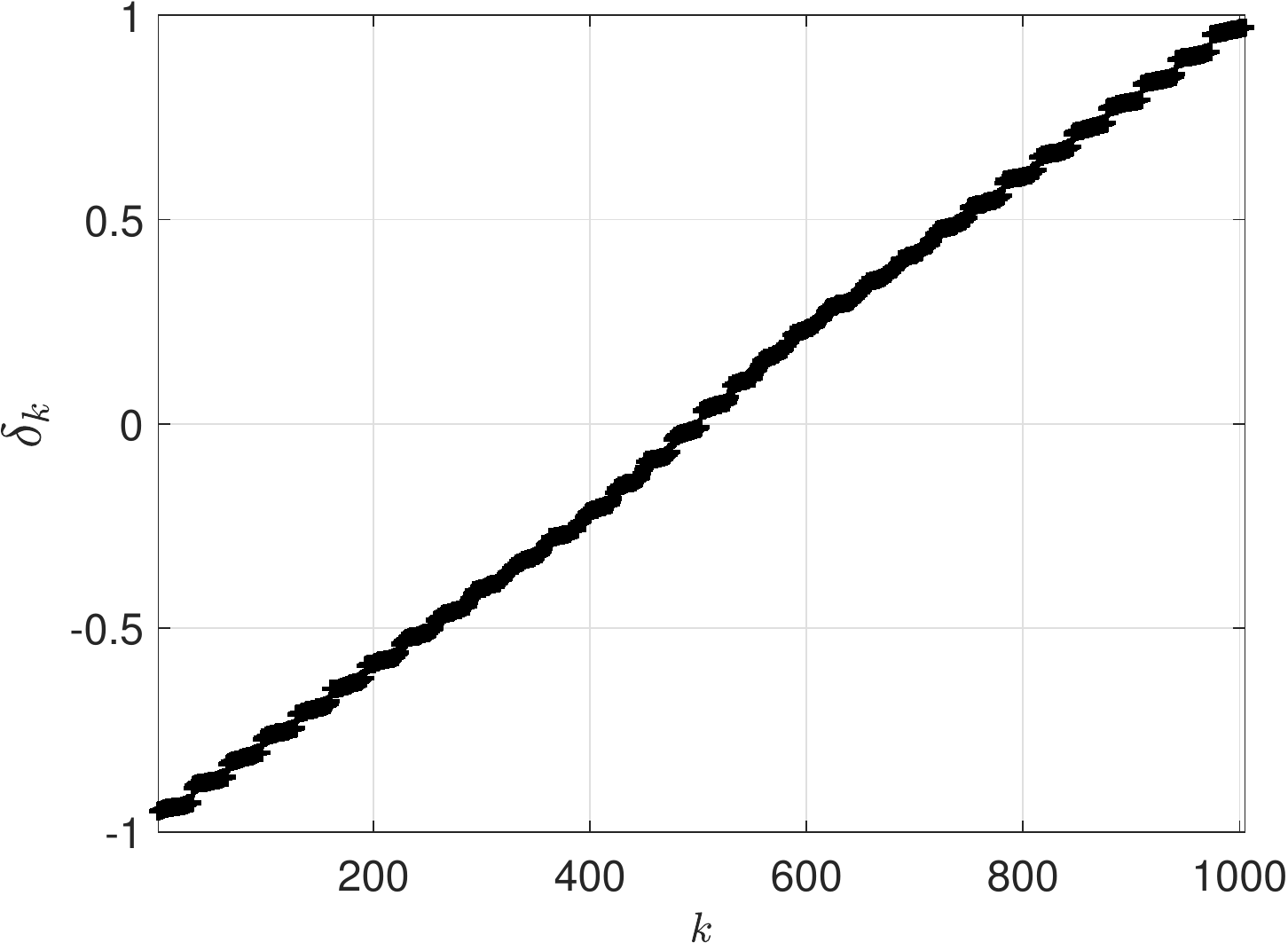}}
	}
	\caption{The (sorted) computed values of the real constants $\delta_1,\delta_2,\ldots,\delta_m$ for $m=4$ (top, left),  $m=10$ (top, right),  $m=123$ (bottom, left), and for  $m=1005$ (bottom, right).}
	\label{fig:delta}
\end{figure}

For the four examples, we plot in Figure~\ref{fig:delta} the computed values of the real constants $\delta_1,\delta_2,\ldots,\delta_m$ (sorted from the smallest to the largest). We see from the figures that the sorted values of these constants almost lie on a straight line for large $m$. 

These examples demonstrate the computational effectiveness of the developed method of integral equations. We select a particular type of 2D structures, namely, a square with a large square hole. Such a hole models a cavity or a defect. The slits model perfectly conducting nanotubes randomly distributed in the bulk medium. The number of slits increasing from $4$ to $1000$ clearly display the local fields in the considered composites. Different inclinations of slits (nanotubes) in Figure \ref{fig:4slits} show that the slit perpendicular to the external flux does not essentially disturb it contrary to the slit parallel to the external flux. The flux intensity can increase in 2 times due to the nanotube. Such an irregularity of the flux was noted in \cite{Ryl, nano3} for analogous problems. With increasing the number of slits the perturbations of the local fields become smaller and a heterogeneous structure can be homogenized. The same examples demonstrate the elastic stress perturbations in the anti-plane problem. For elasticity problem, the modulus of the flux $q$ has to be replaced by the stress intensity $\sqrt{\sigma_{21}^2+\sigma_{31}^2}$ where $\sigma_{ij}$ stands for the components of the stress tensor.

\begin{table}[ht]
	\centering
	\begin{tabular}{@{}cccc@{}}\toprule
		Number of rigid	& Side length of  &  \multirow{ 2}{*}{$f(\eta(t))$} & \multirow{ 2}{*}{$U$ and $q$}  \\ 
		line inclusions & the inner square & \\ \midrule
		4     & 0.5 & 6.69      & 46.76  \\
		10    & 0.4 & 26.41     & 63.10  \\
		123   & 0.3 & 209.03    & 63.10  \\
		1005  & 0.1 & 1989.21   & 90.7   \\ \bottomrule
	\end{tabular}
	\caption{The total CPU time in seconds required to compute the boundary values of the complex distribution temperature $f$ in $w$-plane; and the values of the temperature distribution $U$ and the heat flux $q$.}
	\label{tab:time}%
\end{table}


\begin{figure}[ht] %
	\centerline{
		\scalebox{0.45}{\includegraphics[trim=1cm 0 1cm 0,clip]{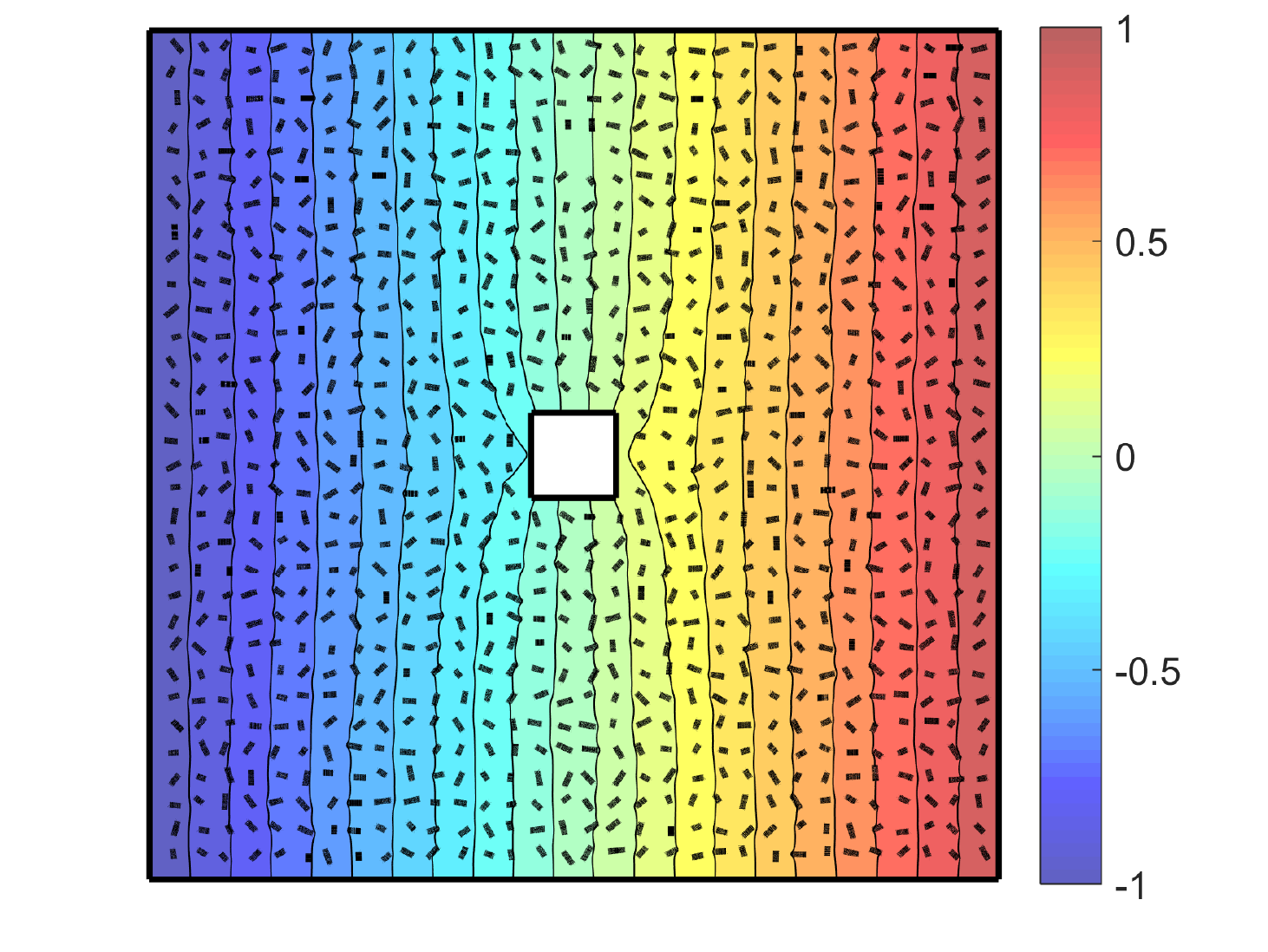}}
		\hfill
		\scalebox{0.45}{\includegraphics[trim=1cm 0 1cm 0,clip]{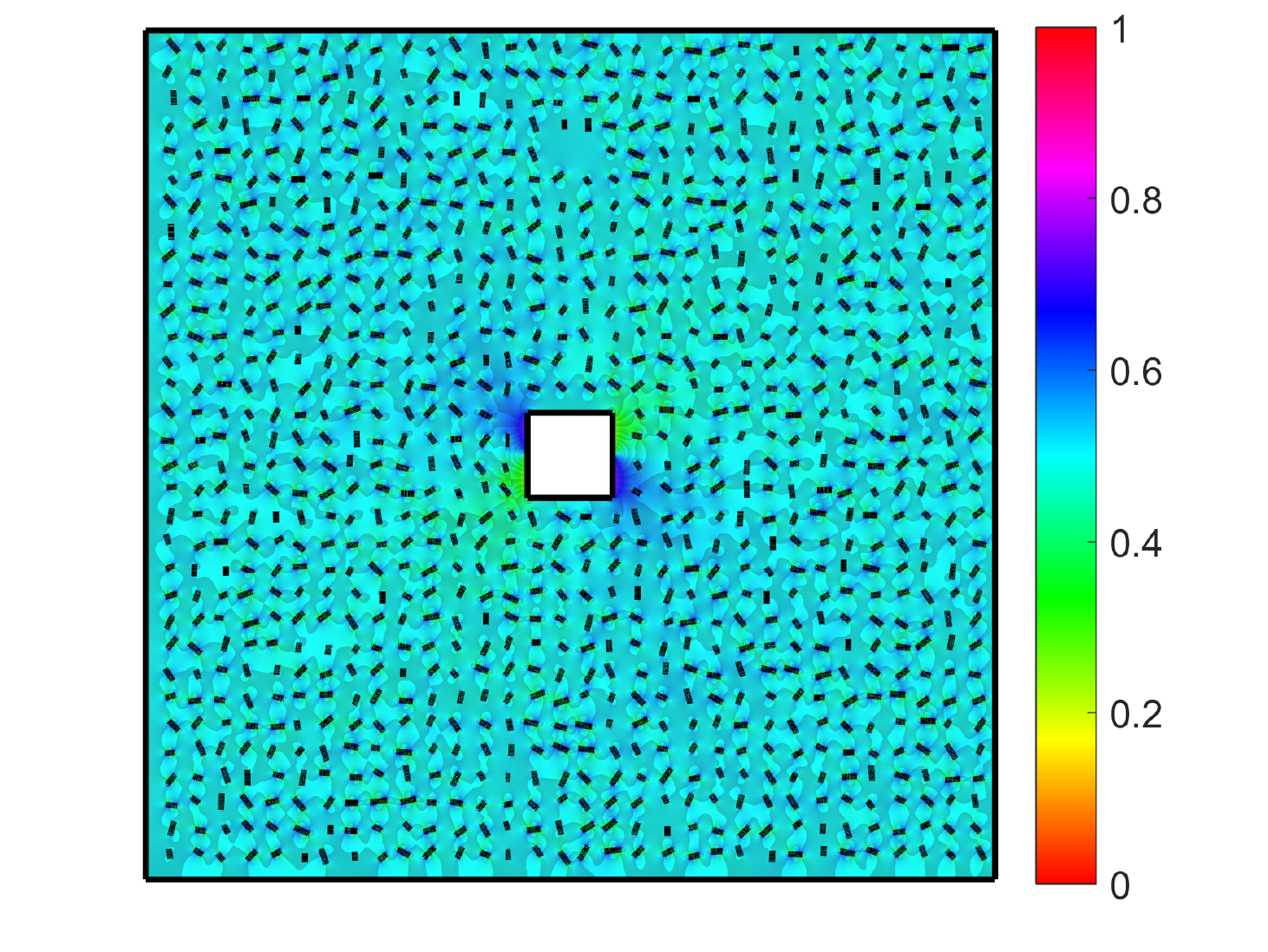}}
	}
	\caption{A contour plot of the temperature distribution $U$ (left) and a phase portrait of the heat flux $q$ (right) for $1005$ CNTs with $r=0.1$.}
	\label{fig:1005slits}
\end{figure}

\section*{Acknowledgments}

We would like to thank Vladimir Mityushev for suggesting this research problem and for his helpful comments and discussions.


\begin{thebibliography}{10}
	
	
	\bibitem{nano1}
	S.M. Lebedev, O.S. Gefle, E.T. Amitov, E.S. Nin, A.E. Bezrodny, M.R. Predtechenskiy,
	Conductive carbon nanotube-reinforced polymer composites and their characterization,  IEEE Transactions on Dielectrics and Electrical Insulation Volume: 23 , Issue: 3 , June 2016
	
	\bibitem{nano2}
	Robert J. Young, Ian A. Kinloch, Lei Gong, Kostya S. Novoselov, The mechanics of graphene nanocomposites: A review. Composites Science and Technology 72 (2012) 1459-1476. 
	
	\bibitem{nas-bvp}
	S.A.A.~{Al-Hatemi}, A.H.M.~{Murid} and M.M.S.~{Nasser}.
	\newblock {A boundary integral equation with the generalized Neumann kernel for a mixed boundary value problem in unbounded multiply connected regions}.
	\newblock {\em Bound. Value Probl.}, Article number: 54, 2013.
	
	\bibitem{Atk97}
	K.E.~{Atkinson}.
	\newblock {The Numerical Solution of Integral Equations of the Second Kind}.
	\newblock Cambridge University Press, Cambridge, 1997.
	
	\bibitem{dry}
	P.~{Dryga\'s}.
	\newblock {New approach to mathematical model of elastic in two-dimensional
		composites}.
	\newblock In P.~{Dryga\'s} and S.~{Rogosin}, editors, {\em
		Modern Problems in Applied Analysis}, pages 87--100. Birkh\"auser, 2018.
	
	\bibitem{gak}
	F.D.~{Gakhov}.
	\newblock {Boundary Value Problems}.
	\newblock Pergamon Press, Oxford, 1966.
	
	
	\bibitem{Gre-Gim12}
	L.~{Greengard} and Z.~{Gimbutas}.
	\newblock {{FMMLIB2D}: A {MATLAB} toolbox for fast multipole method in two dimensions}.
	\newblock Version 1.2, \url{http://www.cims.nyu.edu/cmcl/fmm2dlib/fmm2dlib.html}. Accessed 1 Jan 2018.
	
	\bibitem{Nas-log}
	J.~{Liesen}, O.~{S\'ete} and M.M.S.~{Nasser}.
	\newblock {A fast and accurate computation of the logarithmic capacity of compact sets}. 
	\newblock {\em Comput. Methods Funct. Theory},  17:689--713, 2017.
	
	\bibitem{Mit-chap}
	V.~{Mityushev}.
	\newblock {Mixed problem for Laplace's equation in an arbitrary circular multiply connected domain}.
	\newblock In P.~{Dryga\'s} and S.~{Rogosin}, editors, {\em
		Modern Problems in Applied Analysis}, pages 135--152. Birkh\"auser, 2018.
	
	\bibitem{Mit-Rog}
	V.V.~{Mityushev} and S.V.~{Rogosin}.
	\newblock {Constructive Methods to Linear and Non-linear Boundary Value Problems of the Analytic Function}.
	\newblock Chapman \& Hall/CRC, Boca Raton, 2000.
	
	\bibitem{mus}
	N.I.~{Muskhelishvili}.
	\newblock {Some Basic Problems of the Mathematical Theory of Elasticity}.
	\newblock Springer-Science+Business Media, Dordrecht, 1977.
	
	\bibitem{Nas-jmaa11}
	M.M.S.~{Nasser}.
	\newblock {Numerical conformal mapping of multiply connected regions onto the second, third and fourth categories of Koebe's canonical slit domains}.
	\newblock {\em J. Math. Anal. Appl.}, 382:47--56, 2011.
	
	\bibitem{Nas-jsc19}
	M.M.S.~{Nasser}.
	\newblock {Numerical computing of preimage domains for bounded multiply connected slit domains}.
	\newblock {\em J. Sci. Comput.}, 78:582--606, 2019.
	
	\bibitem{Nas-ETNA}
	M.M.S.~{Nasser}.
	\newblock {Fast solution of boundary integral equations with the generalized Neumann kernel}.
	\newblock {\em Electron. Trans. Numer. Anal.}, 44:189--229, 2015.
	
	\bibitem{Nas-Gre}
	M.M.S.~{Nasser} and C.C.~{Green}.
	\newblock {A fast numerical method for ideal fluid flow in domains with multiple stirrers}.
	\newblock {\em Nonlinearity}, 31:815--837, 2018.
	
	\bibitem{Nas-amc}
	M.M.S.~{Nasser}, A.H.M.~{Murid}, M.~{Ismail} and E.M.A.~{Alejaily}.
	\newblock {Boundary integral equations with the generalized {N}eumann kernel for {L}aplace's equation in multiply connected regions}.
	\newblock {\em Appl. Math. Comput.}, 217:4710--4727, 2011.
	
	\bibitem{Nish}
	N.~{Nishimura} and Y.J.~{Liu}.
	\newblock {Thermal analysis of carbon-nanotube composites using a rigid-line inclusion model by the boundary integral equation method}.
	\newblock {\em Comput. Mech.}, 35:1--10, 2004.
	
	\bibitem{Mit-mod}
	E.~{Pesetskaya}, R.~{Czapla} and V.~{Mityushev}.
	\newblock {An analytical formula for the effective conductivity of 2D domains with cracks of high density}.
	\newblock {\em Appl. Math. Model.}, 53:214--222, 2018.
	
	\bibitem{Ryl}
	N.~{Rylko}. 
	\newblock {Edge effects for heat flux in fibrous composites}.
	\newblock {\em Comput. Math. Appl.}, 70:2283--2291, 2015.
	
	\bibitem{Tre-Trap}
	L.N.~{Trefethen} and J.A.C.~{Weideman}.
	\newblock {The exponentially convergent trapezoidal rule}.
	\newblock {\em SIAM Review}, 56:385--458, 2014.
	
	\bibitem{Weg92}
	E.~{Wegert}.
	\newblock {Visual complex functions. An introduction with phase portraits}.
	\newblock Birkh\"auser/Springer Basel AG, Basel, 2012.
	
	\bibitem{Weg-Nas}
	R.~{Wegmann} and M.M.S.~{Nasser}.
	\newblock {The Riemann-Hilbert problem and the generalized Neumann kernel on multiply connected regions}.
	\newblock {\em J. Comput. Appl. Math.}, 214:36--57, 2008.
	
	\bibitem{Wen92}
	G.C.~{Wen}.
	\newblock {Conformal Mapping and Boundary Value Problems}.
	\newblock Amer. Math. Soc., Providence, 1992.
	
	\bibitem{Nas-proc}
	A.A.M.~{Yunus}, A.H.M.~{Murid} and M.M.S.~{Nasser}.
	\newblock {Numerical conformal mapping and its inverse of unbounded multiply connected regions onto logarithmic spiral slit regions and straight slit regions}.
	\newblock {\em Proc. R. Soc. A} 470:20130514, 2013.
	
	\bibitem{Zha04}
	J.~{Zhang}, M.~{Tanaka} and T.~{Matsumoto}.
	\newblock A simplified approach for heat conduction analysis of CNT-based nano-composites.
	\newblock {\em Comput. Meth. Appl. Mech. Eng.} 192:5597--5609, 2004.
	
	\bibitem{nano3}
	N.~{Rylko}. 
	Fractal local fields in random composites. Computers \& Mathematics with Applications 69(3): 247-254 (2015)
	
\end{thebibliography}
\end{document}